\documentstyle[epsf,epsfig,12pt]{article}
\usepackage{cite}
\def\nll{ \nonumber \\}
\def\lb{\left(}
\def\rb{\right)}

\def\z0{Z}
\def\gf{G_{\mu}}
\def\zm{M_{_Z}}

\def\gev{{\hbox{GeV}}}

\def\hm{M_{_H}}
\def\wm{M_{_W}}

\def\barf{\overline f}

\def\barb{\overline b}
\def\bard{\overline d}

\def\barc{\overline c}
\def\bars{\overline s}
\def\baru{\overline u}

\def\barnu{\overline{\nu}}

\def\i3f{I^{(3)}_f}

\def\osp2{16\,\pi^2}
\def\ap2{\left(p^2\right)}

\def\gev{{\hbox{GeV}}}

\def\s0h{\sigma^h_0}
\def\ba{\begin{eqnarray}}
\def\ea{\end{eqnarray}}
\def\nl{\nonumber \\}

\def\beq{\begin{equation}}
\def\eeq{\end{equation}}
\def\bea{\begin{eqnarray}}
\def\eea{\end{eqnarray}}
\def\barr{\begin{array}}
\def\earr{\end{array}}
\def\bc{\begin{center}}
\def\ec{\end{center}}
\def\btab{\begin{tabular}}
\def\etab{\end{tabular}}

\begin{document}

\title{\bf STANDARD MODEL AND ELECTROWEAK INTERACTION: PHENOMENOLOGY\thanks{
Talk given at the 3rd International Symposium on Radiative Corrections,
Cracow, Poland, August 1-5, 1996}
}
\vskip 2cm
\author{
Giampiero PASSARINO$^{ab}$,
}

\date{}

\maketitle

\begin{itemize}

\item[$^a$]
             Dipartimento di Fisica Teorica,
             Universit\`a di Torino, Torino, Italy
\item[$^b$]
             INFN, Sezione di Torino, Torino, Italy

\end{itemize}

\noindent
email:
\\
giampiero@to.infn.it

\vspace{2cm}
The status of the LEP~1 results, the LEP~2 perspectives and some recent
developments in high energy physics at $e^+e^-$ machines are briefly discussed.

\newpage

In the phenomenology of the standard model we have three different phases:
yesterday, i.e. bounds on standard or new physics from virtual effects, today, 
i.e. direct searches for Higgs particles and new physics and tomorrow, which 
hopefully will be testing the intimate gauge structure of the theory.
If we ask ourselves what has been done since '77 we find that

\begin{itemize}

\item[1977-?] The Born-like structure of the MSM, dressed with effective
couplings which contain the potentially large radiative corrections has been
confronted with the available experimental data, resulting in

\vskip 1cm

\begin{enumerate}

\item Stringent bounds on $m_t$ -- Top quark discovered.

\item Loose bounds on $\hm$ -- Higgs boson(s) to be discovered.

\end{enumerate}

But the times they are a-changin'. Now $m_t$ from CDF+D0 is an input 
parameter in the fits for high precision physics, thus the new frontier is 
the Higgs boson mass, with or without new physics. At the same time the chief 
questions remain unaltered, do the experimental data accept the MSM? Are we 
setting sail for the land beyond the edge of the world, where new physics roam?

\end{itemize}

To summarize we present few facts strictly separated from opinions 
by using the most recent experimental data as presented at
ICHEP '96~\cite{ichep}, only few hours ago.
As a prelude we want to show a fit to the Higgs boson mass 
with the uncertainty due to the error on $\alpha(\zm)$ and $m_b$ fully
propagated in the theoretical part of the $\chi^2$ (TOPAZ0~\cite{topaz0}, see 
also BHM~\cite{bhm} and ZFITTER~\cite{zfitter}).
As a result of the fit we find

\begin{eqnarray}
\hm &=& 143.5\,\gev  \nll
\hm &\leq& 431\,\gev \qquad {\hbox{at}} \qquad 95\% \quad {\hbox{CL}}.
\end{eqnarray}

\noindent 
Our results at the minimum of the $\chi^2$ are shown in table~\ref{tab1}

\begin{table}[hbtp]
\begin{center}
\begin{tabular}{|c|c|c|c|}
\hline
O  & Exp. & Theory & Comments\\
\hline
   &      &        & \\
$M_{_H}\,$(GeV) &  --  & $143.5$ (fixed)   &  $< 431$  at $95\%$ CL    \\
   &      &        & \\
$\chi^2$ & &  $18.22/13$ & \\
   &      &        & \\
$m_t\,$(GeV)    &  $175 \pm 6$ & $172 \pm 5$ & penalty 
in the fit \\
   &      &        & \\
$\alpha^{-1}_{light}(M_{_Z})$  &  $128.89 \pm 0.09$ & $128.905 \pm 0.087$ 
&                \\
   &      &        & \\
$\alpha_s(M_{_Z})$  &  --  & $0.1194 \pm 0.0037$ 
& th. err. not included \\
   &      &        & \\
$m_b\,$(GeV) & $4.7 \pm 0.2$ & $4.68 \pm 0.24$  &  \\
   &      &        & \\
$R_l$ & $20.778 \pm 0.029$ & $20.754 \pm 0.025$ & \\
   &      &        & \\
$\sin^2\theta^e_{eff}$ & $0.23061 \pm 0.00047$ & $0.23159 \pm 0.00022$
 & \\
   &      &        & \\
$R_b$ & $0.2178 \pm 0.001144$ & $0.2158 \pm 0.0002$ & correlated \\
   &      &        & \\
$R_c$ & $0.1715 \pm 0.005594$ & $0.1723 \pm 0.0001$ &     "      \\
   &      &        & \\
$\wm\,$(GeV)   & $80.356 \pm 0.125$  & $80.350 \pm 0.031$  & \\
   &      &        & \\
$A^0_{_{FB}}(b)$ & $0.0979 \pm 0.0023$ & $0.1026 \pm 0.0012$ & \\
   &      &        & \\
   &      &        & \\
\hline
\end{tabular}
\end{center}
\label{tab1}
\caption{Theory versus Experiments around the $Z$ resonance.}
\end{table}

\noindent
In this type of fits -- expecially in a recent past -- the main problems have 
been to understand

\begin{itemize}

\item when the $\chi^2(\hm)$ shape is unstable with respect to
normal fluctuations of the experimental data in the large $\hm$ tail .
The question of the stability of the high-$\hm$ tail in the $\chi^2$
must be constantly addressed. 

\item That very stringent bounds on $\hm$ ($\ll 500\,$GeV) were more a
symptom of the clash between LEP and SLD.

\item That $\chi^2_{min}(\hm)$ has an unnatural tendency to be in the forbidden 
region, thus requiring the unnatural introduction of yet another penalty
function.

\item The effect of including the theoretical error in the fit~\cite{yr}

\end{itemize}

\noindent
With the new data the situation looks considerably better. For the first time 
in the LEP~1 history the $\chi^2_{min}(\hm)$ is in the allowed region, the
quality of the fit is more satisfactory than ever, the minimal standard
model and the minimal supersymmetric standard model seem to be on -- more
or less -- equal footing in describing the data. In conclusion however
it is still premature to give something more precise than an approximate
upper bound of $\approx 500\,$GeV at $95\%$ of CL.

Today we have a new perspective anyway. Even though the central value for
$hm$ has become a little higher we have a non negligible probability of
a Higgs boson within the energy range of LEP~2.
Thus we should stop worrying about Tails$\&$Fits and start to understand 
how a Higgs boson - in the LEP~2 range - looks like in a real environment.

Since we have learned how to deal with unstable particles 
in a field theoretical context~\cite{up} then the properties of the Higgs 
boson at LEP~2 must be inferred from the complete analysis of the following 
processes:

\begin{eqnarray}
e^+e^- &\to& \barb b \mu^+\mu^-, \qquad
\barb b e^+e^-, \nl
{}&& \barb b \barnu \nu, \qquad
\barb b \baru u,(\barc c), \nl
{}&& \barb b \bard d,(\bars s), \qquad
\barb b \barb b
\end{eqnarray}

\noindent
with all the complications arising from flavor mis-identification. Thus the
three typical signatures are

\begin{itemize}

\item two jets + a charged lepton pair,
two jets + missing energy and momentum,
four jets

\end{itemize}

The ideal procedure would be to analyze all channels through some event
generator -- the ultimate one -- which should account for the experimental 
setup, include a self-consistent set of radiative corrections and
be interfaced with some standard hadronization package.
A broad separation can be set between the two alternative approaches:

\begin{itemize}

\item Dedicated electroweak calculations as described by CompHEP~\cite{cod1}, 
EXCALIBUR~\cite{cod2}, GENTLE/4fan~\cite{cod3}, HIGGSPV~\cite{cod4}, 
WPHACT~\cite{cod5}, WTO~\cite{cod6}, $\dots$

\item General purpose simulations (PHYTIA~\cite{cod7}, $\dots$)

\end{itemize}

\noindent
Here we would like to present the point of view of a dedicated EW 
calculation performed with WTO and which is aimed to discuss the Higgs boson 
properties by including all diagrams at the $0.1\%$ level of technical 
precision (or better) with the best available set of radiative corrections, i.e.

\begin{enumerate}

\item Initial State QED radiation through the structure function 
approach~\cite{sf},

\item Running quark masses~\cite{rqm},

\item Naive QCD (NQCD) final state corrections,

\end{enumerate}

and with a simulation of some quasi-realistic experimental setup.
This will allow access to all kind of differential distributions and 
to a control over the background.
To this end we have extended the original version of WTO in order to allow 
for the generation of unweigthed events and for the storage of their 
four-momenta. The full description of WTO 2.0 and of the methods adopted to 
generate unweigthed events will be given elsewhere.

The degree of complexity of the calculation is shown in table~\ref{tab2}
by a simple counting of the diagrams which contribute to each channel

\begin{table}[hbtp]
\begin{center}
\begin{tabular}{|c|c|}
\hline
Final state  & Class/max $\#$ of diagrams\\
             &                           \\
\hline
             &                           \\
$\barb b \mu^+\mu^- $  & NC25 \\
$\barb b e^+e^- $  & NC50\\
$\barb b \barnu_{\mu} \nu_{\mu} $  & NC25\\
$\barb b \barnu_e \nu_e $  & NC21\\
$\barb b \baru u,(\barc c) $  & NC33\\
$\barb b \bard d,(\bars s) $  & NC33\\
$\barb b \barb b $  & NC68 \\
             &                           \\
\hline
\end{tabular}
\end{center}
\label{tab2}
\caption{Diagrams required (including gluons)}
\end{table}

Here we briefly summarize the strategy for the calculation as adopted by WTO.
All fermions masses are neglected but for the $b$-quark mass in the
Yukawa coupling and for the $b$-quark, $c$-quark and $\tau$ masses in the
decay width. Quark masses are running and evaluated according to

\begin{eqnarray}
{\bar m}(s) &=& {\bar m}(m^2)\,\exp\left\{
-\,\int_{a_s(m^2)}^{a_s(s)}\,dx \frac{\gamma_m(x)}{\beta(x)}\right\} \;,
\nll
m &=& {\bar m}(m^2)\,\left[ 1 + \frac{4}{3} a_s(m) + K a_s^2(m)\right]\;,
\end{eqnarray}

\noindent
where $m = m_{pole}$ and $K_b \approx 12.4, K_c \approx 13.3$.
The Higgs width is computed with the inclusion of the $H \to gg$ channel. 
The most complete treatment will therefore evolve $\alpha_s$ to the
scale $\mu = \hm$, evaluate the running $b,c$-quark masses and compute

\begin{eqnarray}
\Gamma_{_H} &=& {{G_G\hm}\over {4\,\pi}}\,\left\{ 3\,\left[ 
m_b^2(\hm) + m_c^2(\hm)\right]\,\left[ 1 + 5.67\,{\alpha_s\over \pi}
+ 42.74\,\lb{\alpha_s\over \pi}\rb^2\right]\right.  \nll
{}&+&\left. m_{\tau}^2\right\} + \Gamma_{gg},  \nll
\Gamma_{gg} &=& {{\gf\hm^3}\over {36\,\pi}}\,{{\alpha_s^2}\over {\pi^2}}\,
\lb 1 + 17.91667\,{\alpha_s\over\pi}\rb.
\end{eqnarray}

\noindent
NQCD is included by evoluting $\alpha_s(\wm)$(input) to 
$\alpha_s(\hm)$ and the Higgs boson signal is multiplied by

\begin{equation}
\delta_{_{QCD}} = 1 + 5.67\,{{\alpha_s}\over {\pi}} + 
42.74\,\lb {{\alpha_s}\over {\pi}}\rb^2,  \qquad
\alpha_s = \alpha_s(\hm).
\end{equation}

\noindent
As an example we give in table~\ref{tab3} some of the relevant quantities as
a function of $\hm$.

\begin{table}[hbtp]
\begin{center}
\begin{tabular}{|c|c|c|c|}
\hline
Parameter  & $\hm = 80\,$GeV & $\hm = 90\,$GeV & $\hm = 100\,$GeV\\
\hline
           &                 &                 &                 \\
$\Gamma_{_H}$   & $1.8515\,$MeV  & $2.0601\,$MeV & $2.2734\,$MeV \\
$m_b(\hm)$      & $2.731\,$GeV   & $2.702\,$GeV  & $2.676\,$GeV  \\
$m_c(\hm)$      & $0.553\,$GeV   & $0.547\,$GeV  & $0.542\,$GeV  \\
$\alpha_s(\hm)$ & $0.12557$      & $0.12323$     & $0.12121$     \\
           &                 &                 &                 \\
\hline
\end{tabular}
\end{center}
\label{tab3}
\caption{Input is $\alpha_s(\zm)= 0.123$, $m_b = 4.7\,$GeV and 
$m_c =1.55\,$GeV.}
\end{table}

\noindent
Once we obtain a prediction at a certain level of technical precision still
the question of the theoretical uncertainty remains to be addressed. There are 
several sources for it but no fully reliable estimate of the theoretical error 
can be given, at most we can produce a rough estimate by applying few options 
connected with the choices of the Renormalization Scheme (GENTLE/4fan, WTO). To
illustrate an example of their effect we have evaluated the Higgs background 
at $190\,$GeV and estimated the theoretical error, as shown in table~\ref{tab4}

\begin{table}[hbtp]
\begin{center}
\begin{tabular}{|c|c|}
\hline
Process  & 1-($G_F$ scheme)/($\alpha$ scheme) (permill)\\
\hline
         &                                             \\
$\barb b \barnu_{\mu} \nu_{\mu}$ &  $  0.86 $\\
$\barb b \mu^+ \mu^-           $ &  $  2.23 $\\
$\barb b \nu_e \nu_e           $ &  $  2.44 $\\
$\barb b e^+ e^-               $ &  $  8.05 $\\
$\barb b \baru u               $ &  $ -3.21 $\\
$\barb b \bard d               $ &  $ -3.03 $\\
         &                                             \\
\hline
\end{tabular}
\end{center}
\label{tab4}
\caption{Differences in Renormalization Scheme}
\end{table}

\noindent
Roughly speaking the theoretical uncertainty associated with the choice
of the RS is most severe whenever low-$q^2$ photons dominate (WPHACT, WTO). 
Indeed let us consider the two most popular choices of RS

\begin{itemize}

\item[-] The $\alpha(\zm)$ scheme.

\begin{equation}
s_{_W}^2 = {{\pi\alpha}\over {{\sqrt 2}\gf\wm^2}},  \qquad 
g^2 = {{4\pi\alpha}\over {s_{_W}^2}}
\end{equation}

\item[-] The $G_F$ scheme

\begin{equation}
s_{_W}^2 = 1 - {{\wm^2}\over {\zm^2}}, \qquad
g^2 = 4{\sqrt 2}\gf\wm^2. 
\end{equation}

\end{itemize}

\noindent
In the $G_F$-scheme the e.m. coupling is governed by $\alpha =
1/131.22$ while in the $\alpha(\zm)$-scheme it is $\alpha = 1/128.89$
which accounts for a $2\%$ difference -- about $10\%$ in the two schemes 
at low-$q^2$ for diagrams with two photons.

Other additional sources of uncertainty are in the
Parametrization of the QED structure functions and in the
treatment of the scale in the QCD corrections, expecially so for the 
scale of $\alpha_s$ in the NQCD. The default consists in inserting 
$\alpha_s$(fixed) even for internal gluons. A better choice could be
$\alpha_s$ at the running virtuality (but cuts are required to avoid the
non-perturbative regime).

On top of the theoretical uncertainties there are additional problems, like
flavor mis-identification and
the correct treatment of the background. Experimentally one must extract 
the Higgs signal from all final states consisting of a pair of (imperfectly)
$b$-tagged jets + remaining products -- including the missing ones.
Indeed the probabilities of a light quark, a $c$-quark or a $b$-quark
jet to be confused with a $b$-quark are non zero. The effect of flavor
mis-identification modifies the original branching ratios.
Moreover at LEP~2 a large fraction of Higgs events will be of the type
$\barb b \barnu \nu $ ($\approx 20\%$). There are potentially large 
backgrounds in

\begin{enumerate}

\item $e \nu_e c s$ with flavor mis-identification and the $e$ lost in the
beam-pipe. A safe estimate requires including $m_e$ in the calculation since
we go down to $\theta_e = 0$ where moreover gauge invariance is in danger.

\item $l^+ l^- \barb b$ with the leptons lost in the beam-pipe. Again it 
requires a finite lepton mass because of divergent multi-peripheral diagrams.

\end{enumerate}

\noindent
No reliable estimate has been given so far.

In order to analyze the Higgs signal versus background we fix our set of cuts. 
They are

\begin{enumerate}

\item $M(\barb b) \geq 50\,$GeV, $\mid M(\barf f) - \zm\mid \leq 25\,$GeV.
The former is to suppress the photon mediated $\barb b$ production
-- which decreases for larger $\sqrt{s}$. The latter reduces all contributions
which give a broad $M(\barf f)$ spectrum.

\item Lepton momenta $\geq 10\,$GeV, $E_q \geq 3\,$GeV.

\item Lepton polar angles with the beams $\geq 15^o$.

\item For processes with neutrinos the angle of both b's with the beams
$\geq 20^o$ or of at least one b. Moreover $\theta(l,q) \geq 5^o$.

\end{enumerate}

\begin{figure}[ht]
\vspace{0.1cm}
\centerline{
\epsfig{figure=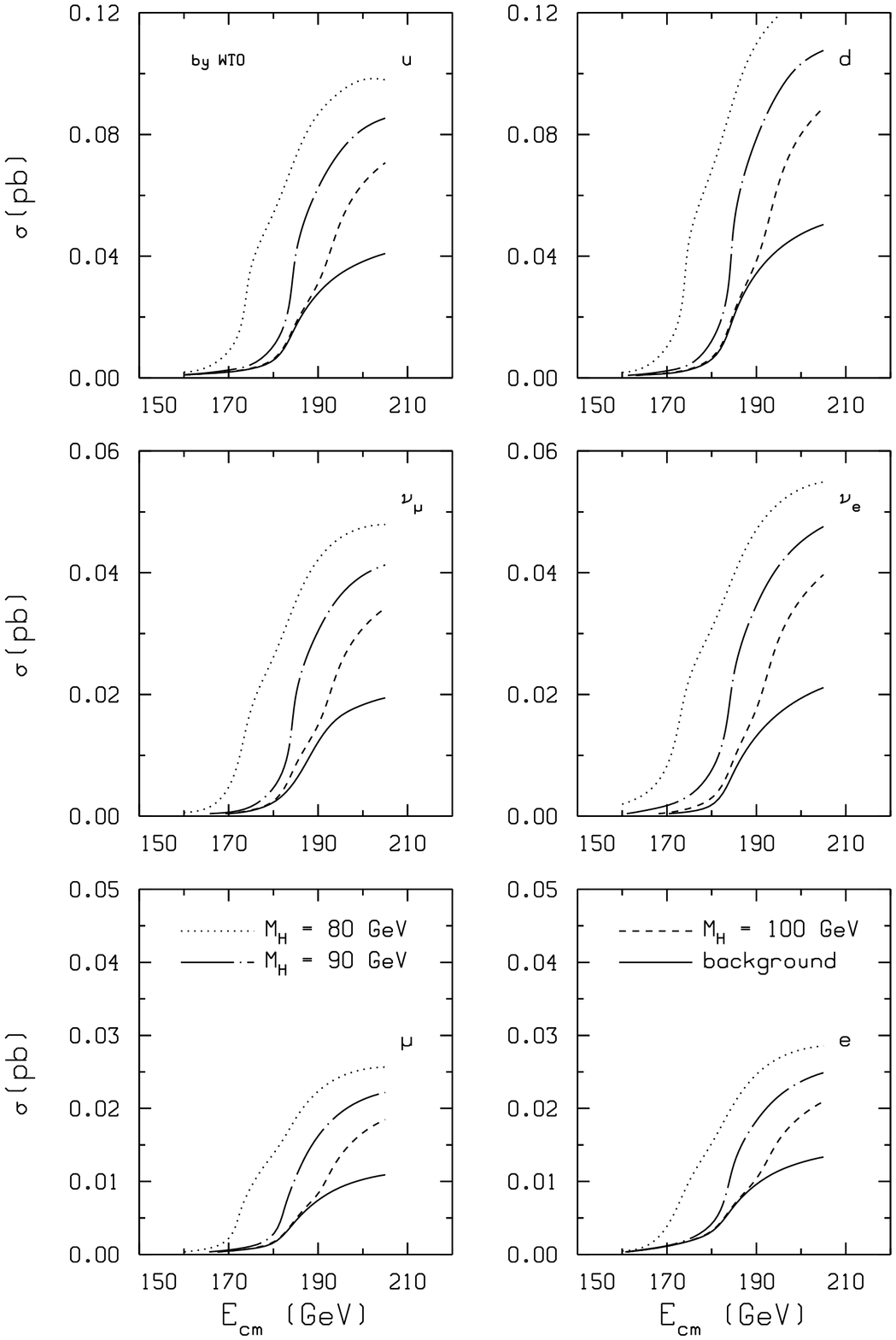,height=20cm,angle=0}
}
\vspace{1.5cm}
\begin{center}
{\bf Fig. 1 Cross sections for $e^+e^- \to \barb b \barf f$}
\end{center}
\label{fig1}
\end{figure}

\noindent
Next we start by showing the cross sections as a function of $\sqrt{s}$ 
for $\hm = 80\,\gev, 90\,\gev, 100\,\gev$ and $\infty$. They are shown in 
Fig.~\ref{fig1}

\noindent
Our findings confirm the rule of thumb $\hm \approx \sqrt{s} - 100\,\gev$ 
for LEP~2 feasibility.
If the Higgs is above $80\,GeV$ the cross section is too small at $\sqrt{s} =
175\,$GeV to allow Higgs discovery, the $\sqrt{s} = 190\,$GeV phase of the 
collider is needed.
At $\sqrt{s} = 190\,$GeV the $ZZ$ background is not negligible
and here is where a dedicated EW calculation becomes useful.

There are several distributions which are of some relevance in the 
Higgs study. They provide informations useful for choosing cuts in the
Higgs searches. Among them we have selected:

\begin{enumerate}

\item The $M(\barb b)$ distribution for all channels but $\barb b \barb b$.
It is useful whenever the direct reconstruction of the invariant mass from
the jets in the process is viable.

\item The missing mass recoil. A knowledge of $\sqrt{s}$ and of the leptonic
final states is required

\begin{equation}
M_{miss}^2 = s - 2\,\sqrt{s}\left(E_{l^+} + E_{l^-}\right) + M^2(l^+l^-)
\end{equation}

\item The visible energy in $\barb b\barnu \nu$, or in general $E(\barb) +
E(b)$. The $b$-quark pairs from the Higgs decay have a sharp peak in the
energy distribution due to the small Higgs width.

\item Angular distributions for the $b$-quark and/or the $\barb$-quark.
In particular the $\cos\theta(\barb b)$ distribution of the total $3$-momentum
${\vec p}_{\barb b}$. However, in general, the signal angular distributions are
very isotropic.

\end{enumerate}

\begin{figure}[ht]
\vspace{0.1cm}
\centerline{
\epsfig{figure=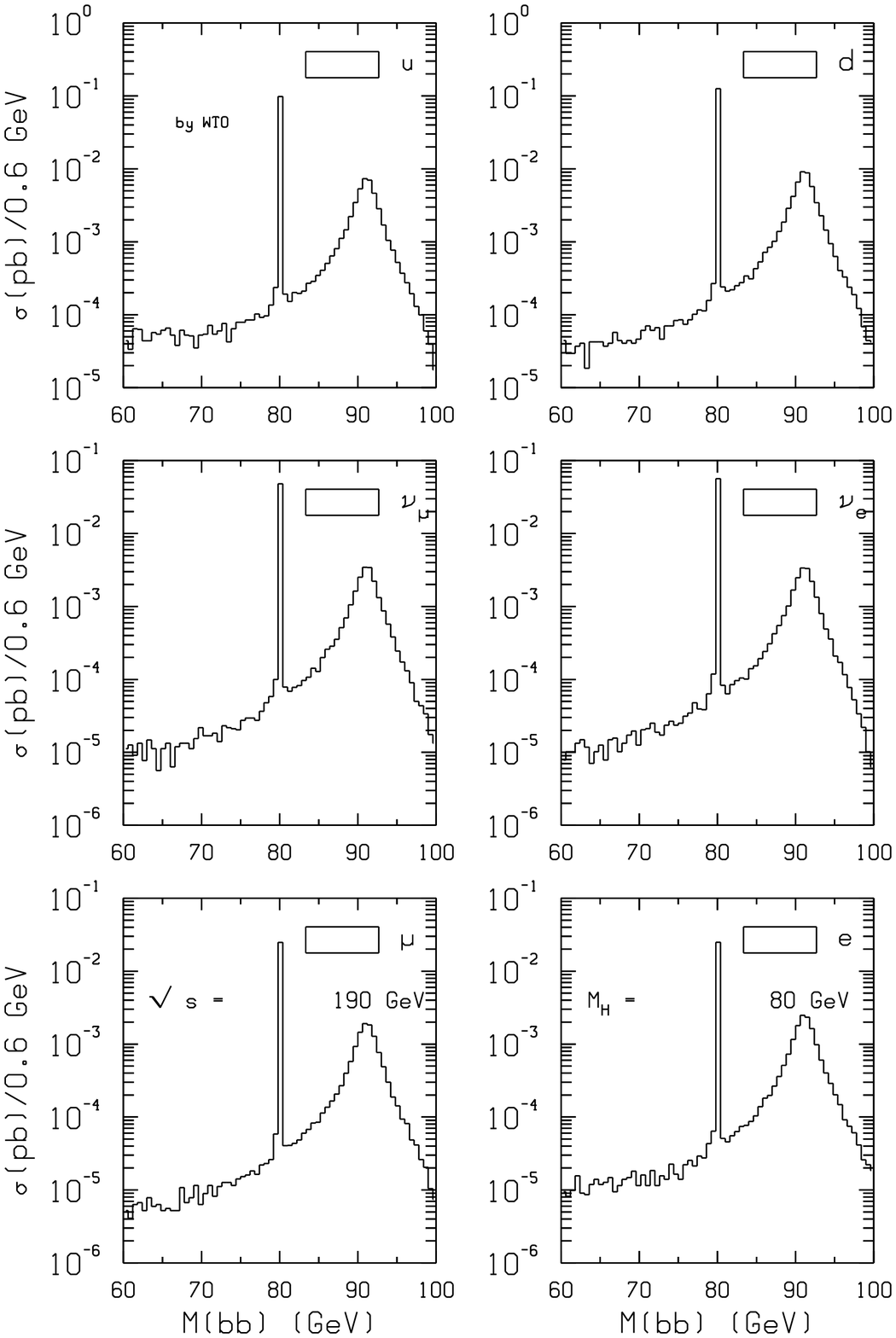,height=20cm,angle=0}
}
\vspace{1.5cm}
\begin{center}
{\bf Fig. 2 The $M(\barb b)$ distribution for $e^+e^- \to \barb b \barf f$.}
\end{center}
\label{fig2}
\end{figure}

\begin{figure}[ht]
\vspace{0.1cm}
\centerline{
\epsfig{figure=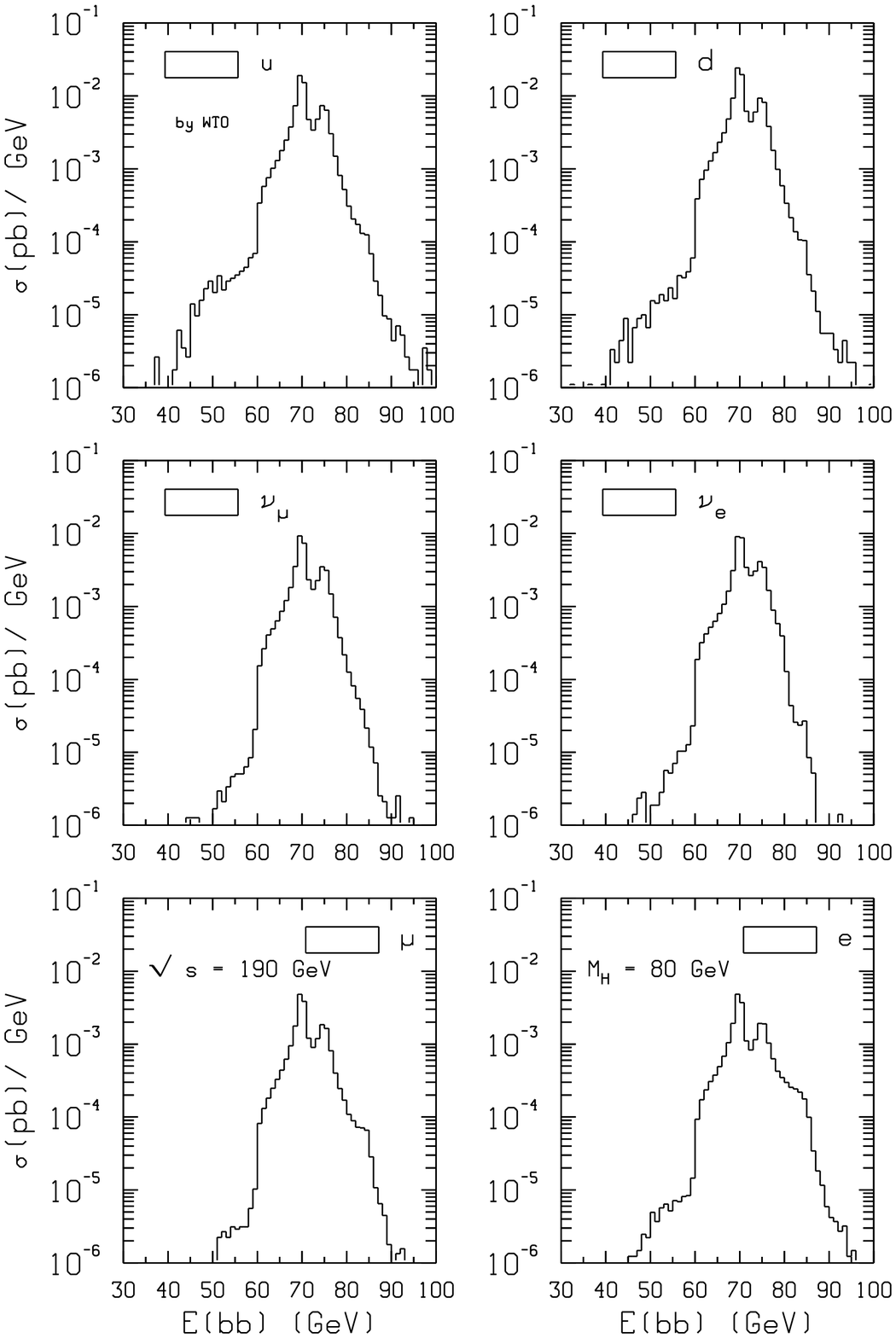,height=20cm,angle=0}
}
\vspace{1.5cm}
\begin{center}
{\bf Fig. 3 The $E(\barb) + E(b)$ distribution for $e^+e^- \to \barb b 
\barf f$.}
\end{center}
\label{fig3}
\end{figure}

\begin{figure}[ht]
\vspace{0.1cm}
\centerline{
\epsfig{figure=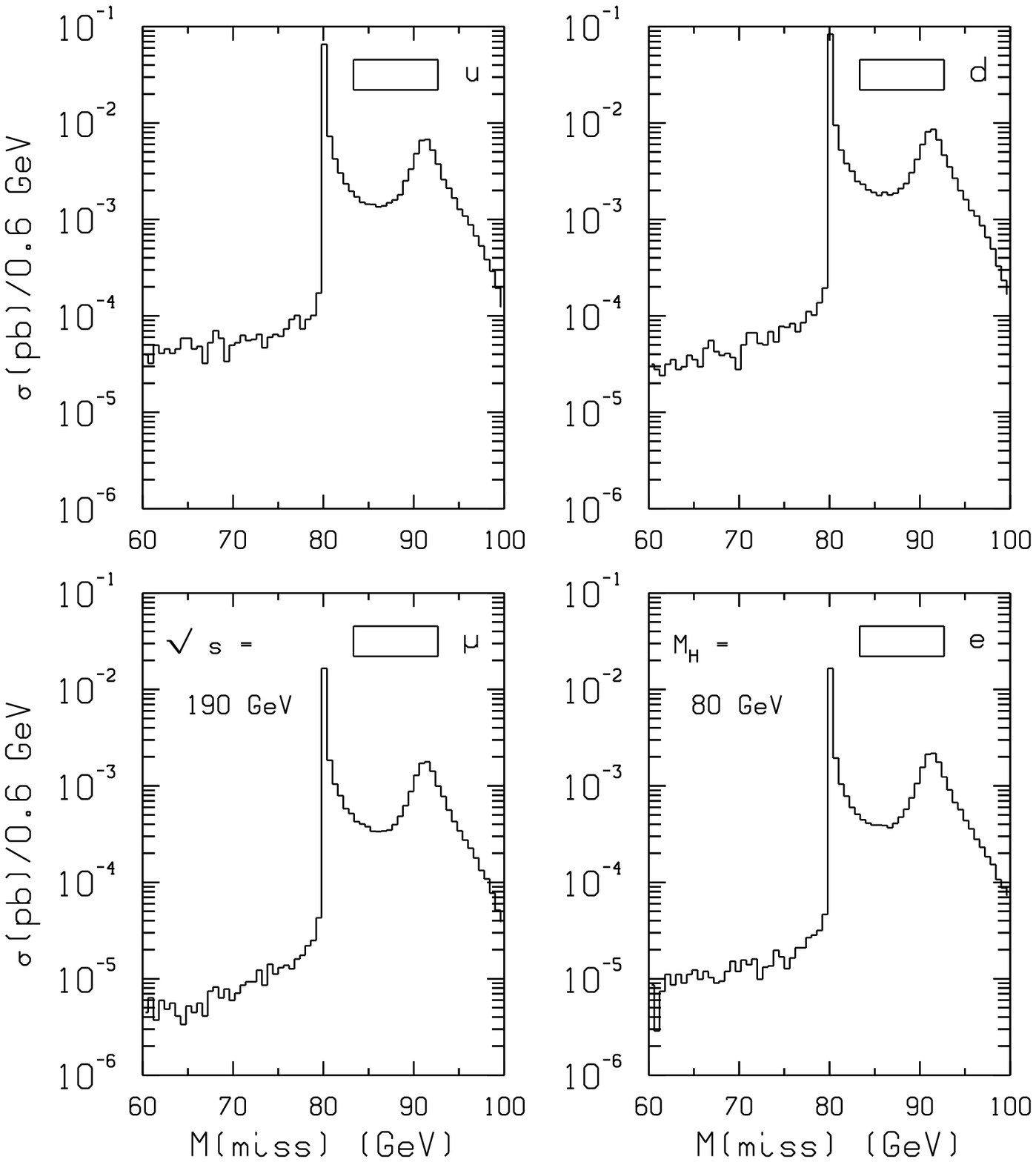,height=20cm,angle=0}
}
\vspace{1.5cm}
\begin{center}
{\bf Fig. 4  The $M_{miss}$ distribution for $e^+e^- \to \barb b
\barf f$.}

\end{center}
\label{fig4}
\end{figure}

\noindent
There is not enough space here to discuss all the distributions and for this 
reason we have limited our presentation to some selected sample where 
$\sqrt{s} = 190\,$GeV and $\hm = 80\,$GeV.
In Fig.~\ref{fig2} we have shown the $M(\barb b)$ distribution while
$E(\barb) + E(b)$ is given in Fig.~\ref{fig3}. Finally the missing mass recoil
is shown in Fig.~\ref{fig4}. 

\noindent
The previous examples are meant to show the quality of the results
achievable with a dedicated EW calculation its technical precision and
the connected theoretical uncertainty.
All diagrams have been included and before hadronization
this is the status of art.

Our final goal will be to discuss the new frontier: $e^+e^-$ annihilation 
beyond LEP~2.
Among the many facets of the physics at  NLC we want to select just
one particular issue~\cite{up}:

\begin{itemize}

\item Any calculation for $e^+e^- \to 4(6)$-fermions is only nominally a 
tree level approximation because of the presence of charged and neutral, 
unstable vector bosons.
Unstable particles require a special care and their propagators, in 
some channels, must necessarily include an imaginary part or in other words 
the corresponding $S$-matrix elements will show poles shifted into the complex 
plane (actually on the second Riemann sheet). 

The introduction of a width into the propagators will inevitably result, 
in some cases, into a breakdown of the relevant Ward identities of the theory 
with a consequent violation of some well understood cancellation mechanism.

\end{itemize}

\noindent
Among the different choices at our disposal we have:

\begin{enumerate}

\item Running width for a vector boson (both charged and neutral) in any
$s$-channel,

\begin{equation}
\Delta^{-1}_{_V}(s) = s - M_{_V}^2 + i\,\frac{s}{M_{_V}}\Gamma_{_V}.
\end{equation}

\noindent
as dictated by Dyson re-summation of the self energy diagrams.

\item Ad hoc fixed width

\begin{equation}
\Delta^{-1}_{_V}(s) = s - M_{_V}^2 + i\,M_{_V}\Gamma_{_V}.
\end{equation}

\item Improved fixed with, as derived by analyzing the solution for the complex
pole. Here one uses a fixed complex mass

\begin{equation}
\mu^2_{_V} = M^2_{_V} - \Gamma^2_{_V} - i\,M_{_V}\Gamma_{_V}\left(1 -
{{\Gamma^2_{_V}}\over {M^2_{_V}}}\right).
\end{equation}

\end{enumerate}

What are the predictions that we can make for an $e^+e^-$ annihilation
into four fermions at large energies? An example for the process
$e^+e^- \to d \baru c \bars$ with QED correction (but without beamsthralung)
at large energies is shown in table~\ref{tab5}

\begin{table}[hbtp]
\begin{center}
\begin{tabular}{|c|c|c|c|}
\hline
$\sqrt{s}$(TeV) & running width & fixed width & improved fixed width \\
\hline
                &               &             &                      \\
0.5 & 0.8651(2) & 0.8612(5)     & 0.8614(5) \\
1   & 0.3650(1) & 0.3505(1 )    & 0.3505(1) \\
1.5 & 0.2267(3) & 0.1953(3)     & 0.1957(4) \\
2   & 0.1827(7) & 0.1279(4)     & 0.1275(2) \\
                &               &             &                      \\
\hline
\end{tabular}
\end{center}
\label{tab5}
\caption{Cross sections for $e^+e^- \to d \baru c \bars$ (No QCD).}
\end{table}

\noindent
From our calculation one can see that at $2\,$TeV running versus fixed width 
(for both $W$ and $Z$) amounts to quite some difference.
The running width prediction is decreasing at a slower rate because
a running width is actually spoiling gauge invariance which, in
turns, means that the underlying unitarity cancellation at asymptotic
energies for $e^+e^- \to W^+W^-$ is not working anymore.

The solution is known by now but let's make a step backward, we have a somehow
related problem already at LEP~2.
A breakdown of the relevant Ward identities of the theory  
in the $e^+e^- \to e^- \barnu_e \nu_{\mu} \mu^+(u \bard)$ case results into 
a numerical catastrophe.
This is already at LEP~2 energies and it has to do with the
interaction of $W$'s with photons and with small scattering angle of the
electron.
The solution is known: one can remain minimalist and include the 
imaginary parts of all other diagrams, the so called Fermion loop scheme.
It goes like this: take $e^+e^- \to e^- \barnu_e u \bard$ and compute
the cross section with $\pi-\theta_{cut} < \theta_{e} < \theta_{cut}$.
We do that for $\sqrt{s} = 175\,$GeV and including QED, but not QCD, and show 
the results in table~\ref{tab6}.

\begin{table}[hbtp]
\begin{center}
\begin{tabular}{|c|c|c|c|}
\hline
$\theta_{cut}$(deg) & running width & FL scheme & fixed width \\
\hline
                     &               &           &             \\
$10^o$  & 0.48681(4) &  0.48692(4)   &   0.48674(4) \\
$5^o$   & 0.49755(5) &  0.49767(4)   &   0.49748(5)  \\
$1^o$   & 0.5184(5)  &  0.5125(4)    &   0.5123(4) \\
$0.5^o$ & 0.546(2)   &  0.5181(6)    &   0.5179(5) \\
                     &               &           &             \\
\hline
\end{tabular}
\end{center}
\label{tab6}
\caption{Restoring gauge invariance at LEP~2}
\end{table}

\noindent
However this is more or less an academical problem, a cut on the electron 
angle will always be imposed and at, say, $5^o$ any choice is equally good.

At NLC restoring the unitarity cancellation requires more: the full ${\cal O}
(\alpha)$ fermionic corrections, both real and imaginary must be included.
This we call Complete Fermion Loop scheme.
All fermionic radiative corrections are organized in terms of:

\begin{itemize}

\item[a] The complex poles $s_{_W}$ and $s_{_Z}$ (on the second Riemann sheet)
for the $W$ and $Z$ vector boson.

\item[b] The running of $\alpha(s)$, the e.m. coupling constant.
The running of $g^2(s)$ the $SU(2)$ coupling constant.

\item[c] The one loop fermionic vertices.

\end{itemize}

There is of course a little problem with experimental input data,
too many: $\zm, \wm, \alpha, G_{_F}$.
Thus a consistency condition can be imposed and the only other parameter
available should be fixed, the top quark mass.
This is, strictly speaking, non completely satisfactory since
QCD corrections have not been applied, not even for the self-energies and
bosonic corrections cannot be neglected in this matters.
Indeed when we include all the available radiative corrections in the on-shell
scheme we find the results of table~\ref{tab7} with a quite remarkable
agreement with the experimental data.

\begin{table}[hbtp]
\begin{center}
\begin{tabular}{|c|c|c|c|c|}
\hline
$\wm/\hm$ (GeV) & 60 & 300 & 1000 & FL \\
\hline
                &    &     &  & \\
80.10 & 119.1 & 137.3 & 154.2 & 104.8\\
80.26 & 148.1 & 165.3 & 181.2 & 132.2\\
80.42 & 174.4 & 190.7 & 206.1 & 157.2\\
                &    &     &  & \\
\hline
\end{tabular}
\end{center}
\label{tab7}
\caption{Calculated $m_t$ for $\alpha_s(\zm) = 0.123$ and different Higgs and 
$W$ masses. The last column is the Fermion Loop prediction.}
\end{table}

\noindent
Instead in the Fermion Loop scheme we obtain the results of the last column
in table~\ref{tab7}, clearly showing the quality of the approximation.

\noindent
Once the relevant Ward identities have been proven the corresponding
cross sections can be computed, giving a field-theoretically and numerically
correct description of $e^+e^-$ annihilation at high energies.
While the theoretical bases, including the whole set of Ward
identities, compact expressions for the vertices, the full machinery
of the running coupling constants is by now fully developed 
the numerical implementation into a realistic code and its debugging is still 
not at anchor (the version of WTO showing items a-d is in progress).

The Fermion Loop scheme, being the first self-consistent example of radiative
corrections for LEP~2 physics and beyond, will eventually open the road for
other (desperately) wanted calculations

\begin{itemize}

\item Complete QCD corrections to CC10 processes,
${\cal O}(\alpha\times{\hbox{const}})$ QED corrections to all
LEP~2 processes and full ${\cal O}(\alpha)$ EW corrections for LEP~2 
processes. 

\end{itemize}

%===============================================================================

\end{document}